\documentstyle[12pt]{article}
\input epsf

\catcode`\@=11
\long\def\@makefntext#1{
\protect\noindent \hbox to 3.2pt {\hskip-.9pt
$^{{\ninerm\@thefnmark}}$\hfil}#1\hfill}		

\def\@makefnmark{\hbox to 0pt{$^{\@thefnmark}$\hss}}  
\def\ps@myheadings{\let\@mkboth\@gobbletwo
\def\@oddhead{\hbox{}
\rightmark\hfil\ninerm\thepage}
\def\@oddfoot{}\def\@evenhead{\ninerm\thepage\hfil
\leftmark\hbox{}}\def\@evenfoot{}
\def\sectionmark##1{}\def\subsectionmark##1{}}

\renewcommand{\thefootnote}{\fnsymbol{footnote}}

\newcounter{sectionc}\newcounter{subsectionc}\newcounter{subsubsectionc}
\renewcommand{\section}[1] {\vspace*{0.6cm}\addtocounter{sectionc}{1}
\setcounter{subsectionc}{0}\setcounter{subsubsectionc}{0}\noindent
	{\normalsize\bf\thesectionc. #1}\par\vspace*{0.4cm}}
\renewcommand{\subsection}[1] {\vspace*{0.6cm}\addtocounter{subsectionc}{1}
	\setcounter{subsubsectionc}{0}\noindent
	{\normalsize\it\thesectionc.\thesubsectionc. #1}\par\vspace*{0.4cm}}
\renewcommand{\subsubsection}[1]
{\vspace*{0.6cm}\addtocounter{subsubsectionc}{1}
	\noindent {\normalsize\rm\thesectionc.\thesubsectionc.\thesubsubsectionc.
	#1}\par\vspace*{0.4cm}}

\newcounter{appendixc}
\newcounter{subappendixc}[appendixc]
\newcounter{subsubappendixc}[subappendixc]

\renewcommand{\appendix}[1] {\vspace*{0.6cm}
        \refstepcounter{appendixc}
        \setcounter{figure}{0}
        \setcounter{table}{0}
        \setcounter{equation}{0}
        \renewcommand{\thefigure}{\Alph{appendixc}.\arabic{figure}}
        \renewcommand{\thetable}{\Alph{appendixc}.\arabic{table}}
        \renewcommand{\theappendixc}{\Alph{appendixc}}
        \renewcommand{\theequation}{\Alph{appendixc}.\arabic{equation}}
        \noindent{\bf Appendix \theappendixc #1}\par\vspace*{0.4cm}}

\def\abstracts#1{{
\centering{\begin{minipage}{12.2truecm}\vspace*{.1cm}
        \footnotesize\baselineskip=12pt\noindent
	\parindent=0pt #1
	\end{minipage}}\par}}

\renewenvironment{thebibliography}[1]
	{\begin{list}{\arabic{enumi}.}
	{\usecounter{enumi}\setlength{\parsep}{0pt}
\setlength{\leftmargin 1.25cm}{\rightmargin 0pt}
	 \setlength{\itemsep}{0pt} \settowidth
	{\labelwidth}{#1.}\sloppy}}{\end{list}}

\newcounter{itemlistc}
\newcounter{romanlistc}
\newcounter{alphlistc}
\newcounter{arabiclistc}

\newcommand{\fcaption}[1]{
        \refstepcounter{figure}
        \setbox\@tempboxa = \hbox{\footnotesize Fig.~\thefigure. #1}
        \ifdim \wd\@tempboxa > 6in
           {\begin{center}
        \parbox{6in}{\footnotesize\baselineskip=12pt Fig.~\thefigure. #1}
            \end{center}}
        \else
             {\begin{center}
             {\footnotesize Fig.~\thefigure. #1}
              \end{center}}
        \fi}

\newcommand{\tcaption}[1]{
        \refstepcounter{table}
        \setbox\@tempboxa = \hbox{\footnotesize Table~\thetable. #1}
        \ifdim \wd\@tempboxa > 6in
           {\begin{center}
        \parbox{6in}{\footnotesize\baselineskip=12pt Table~\thetable. #1}
            \end{center}}
        \else
             {\begin{center}
             {\footnotesize Table~\thetable. #1}
              \end{center}}
        \fi}

\def\@citex[#1]#2{\if@filesw\immediate\write\@auxout
	{\string\citation{#2}}\fi
\def\@citea{}\@cite{\@for\@citeb:=#2\do
	{\@citea\def\@citea{,}\@ifundefined
	{b@\@citeb}{{\bf ?}\@warning
	{Citation `\@citeb' on page \thepage \space undefined}}
	{\csname b@\@citeb\endcsname}}}{#1}}

\newif\if@cghi
\def\cite{\@cghitrue\@ifnextchar [{\@tempswatrue
	\@citex}{\@tempswafalse\@citex[]}}
\def\citelow{\@cghifalse\@ifnextchar [{\@tempswatrue
	\@citex}{\@tempswafalse\@citex[]}}
\def\@cite#1#2{{$\null^{#1}$\if@tempswa\typeout
	{IJCGA warning: optional citation argument
	ignored: `#2'} \fi}}

\font\twelvebf=cmbx10 scaled\magstep 1
 1
 1

\font\ninerm=cmr9

\def\M{{\cal M}}
\def\o{\over}
\def\dr{\mbox{\footnotesize$\overline{\it DR}$}~}
\def\ms{\mbox{\footnotesize$\overline{\it MS}$}~}
\def\roughly#1{\raise.3ex\hbox{$#1$\kern-.75em\lower1ex\hbox{$\sim$}}}

\begin{document}

\hspace*{\fill}JHU-TIPAC-95022\\
\hspace*{\fill}hep-ph/9508219\\
\vspace{.5in}

\centerline{\twelvebf THE STRONG COUPLING AND THE BOTTOM MASS}
\baselineskip=15pt
\centerline{\twelvebf IN SUPERSYMMETRIC GRAND UNIFIED THEORIES
\footnote{Talk given at the PASCOS Symposium/Johns Hopkins Workshop,
Baltimore, MD, March 22-25, 1995.}}
\baselineskip=22pt

\centerline{\footnotesize DAMIEN PIERCE}
\baselineskip=13pt
\centerline{\footnotesize\it Department of Physics and Astronomy}
\centerline{\footnotesize\it Johns Hopkins University}
\centerline{\footnotesize\it 3400 N. Charles Street}
\centerline{\footnotesize\it Baltimore, Maryland 21218, USA}
\centerline{\footnotesize E-mail: pierce@planck.pha.jhu.edu}

\vspace*{0.9cm}
\abstracts{
We apply the full one-loop corrections to the masses, gauge couplings,
and Yukawa couplings in the minimal supersymmetric standard model.
We focus on predictions for the strong coupling and the bottom-quark
pole mass in the context of SU(5) supersymmetric grand unification.
We discuss our results in both the small and large $\tan\beta$ regimes.
We demonstrate that the finite (non-logarithmic)
corrections to the weak mixing angle are essential in determining
$\alpha_s$ and $m_b$ when some superpartner masses are light.
Minimal SU(5) predicts acceptable $\alpha_s$ and $m_b$
only at small $\tan\beta$ with SUSY masses of $\cal O$(TeV).
The missing doublet model accommodates gauge and Yukawa
coupling unification for small or large $\tan\beta$,
and for any SUSY mass scale. In the large
$\tan\beta$ case, the bottom mass is acceptable only if the
Higgsino mass parameter $\mu$ is positive.}
\normalsize\baselineskip=15pt
\setcounter{footnote}{0}
\renewcommand{\thefootnote}{\alph{footnote}}

\section{Introduction}
This talk is organized as follows. First, we briefly
discuss the supersymmetric standard model, in order to introduce
the parameter space we are considering.
Next, we discuss the prediction for the strong coupling constant
and the bottom pole mass with and without including GUT threshold
effects, in the small $\tan\beta$ case. Lastly we discuss our results
in the large $\tan\beta$ regime.

The minimal supersymmetric model is attractive in that it solves the
hierarchy problem. It does this, however,  at the expense of doubling
the number of degrees of freedom of the standard model. This is
problematic, as supersymmetry must be broken and hence the number of
new parameters needed to describe the model is in general quite large.
An organizing principle is needed in order to reduce the number of
parameters. In a minimal supergravity scenario the number of
supersymmetry breaking parameters
needed to describe the supersymmetric model are few: a universal
scalar mass $M_0$, a universal gaugino mass $M_{1/2}$, a universal
trilinear scalar coupling $A_0$, and a bilinear scalar coupling $B$.
In addition there is a supersymmetric Higgs mass term, $\mu$.
Given values for these five parameters at the GUT scale, we use the
renormalization group equations~\cite{RGEs} (RGE's)
to determine the various
particle masses and couplings at the weak scale. For a large top-quark
mass, one of the Higgs boson masses is driven negative, and the
radiative breaking of $SU(2)\times U(1)$ symmetry becomes manifest.
The Higgs bosons obtain vev's $v_1$ and $v_2$, and we can determine
the $Z$-boson mass. In practice it is convenient to assume
electroweak symmetry breaks radiatively and to take $M_Z$ as an input
parameter, as well as the ratio of vev's $\tan\beta\equiv v_2/v_1$.
We then determine $\mu^2$ and $B$ from the symmetry breaking conditions.
To summarize, then, the supersymmetric model is parametrized by
$$M_0,\ \ M_{1/2},\ \ A_0,\ \ \tan\beta,\ \ {\rm sign}(\mu).$$

The exact one-loop corrections to the masses, gauge couplings and
Yukawa couplings of the minimal supersymmetric model are described
in Ref. [2].  These corrections are essential ingredients
for accurate tests of grand unification.  They allow one to extract
the underlying \dr parameters from a given set of measured observables.
The \dr parameters can then be run up to a high scale to explore
the consequences of different unification hypotheses.

Alternatively, the radiative corrections can be used to translate
various limits into excluded regions of the \dr parameter  space.  This
is illustrated in Fig.~1, where we show the excluded region of the $M_0,
\ M_{1/2}$ parameter space at the one-loop level, from current
experimental constraints.

\begin{figure}[t]
\epsfysize=3in
\epsffile[30 495 600 755]{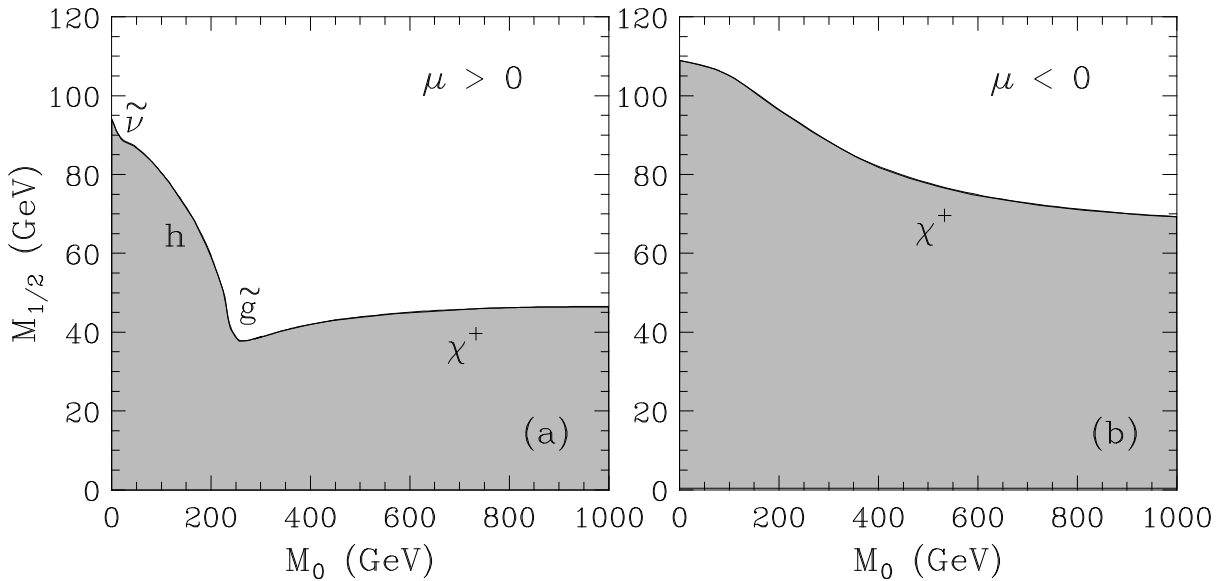}
\begin{center}
\parbox{5.5in}{
\caption[]{\footnotesize Excluded region (shaded)
of the $M_0,\ M_{1/2}$ plane,
for $\tan\beta=2,\ m_t=175$ GeV, $A_0=0$, and $\alpha_s=0.117$.
All masses are evaluated at one-loop.
The symbols indicate which experimental constraint is relevant:
$\chi^+\Rightarrow m_{\chi^+} > 47$ GeV; $\tilde g\Rightarrow
m_{\tilde g}>125$ GeV; $\tilde\nu\Rightarrow m_{\tilde\nu}>42$ GeV;
$h\Rightarrow m_h>60$ GeV.}}
\end{center}
\end{figure}

In the following, we treat all supersymmetric threshold corrections
in a complete one-loop analysis.\footnote{See Chankowski
et al.~\cite{Chankowski} for a similar
treatment of finite corrections to $\sin^2\theta_W$.}  Our work stands
in contrast to most previous studies, which are based on the ``leading
logarithm approximation."  For the gauge coupling threshold corrections,
this approximation involves taking the
standard-model value of $\sin^2\theta_W$ and adding the logarithmic
parts of the SUSY threshold corrections.  The approximation works well
if all of the SUSY particle masses are much greater than $M_Z$, in which
case the decoupling theorem implies that the finite effects of the
SUSY particles are negligible for all low-energy observables.

However, in realistic models it is not unusual for the supersymmetric
spectrum to contain light particles of order the $Z$-mass.  In this
case the leading logarithm approximation breaks down.
This is illustrated in Fig.~2, where we compare the value of
$\alpha_s$ and $m_b$ in the leading logarithm approximation (LLA) with
the value obtained in the full calculation.
In this talk we use the full set of one-loop radiative
corrections to evaluate the \dr gauge and Yukawa couplings.  The \dr
couplings serve as the boundary conditions for the two-loop gauge and
Yukawa coupling renormalization group equations, which
determine the couplings at very high scales.
\begin{figure}[t]
\epsfysize=3in
\epsffile[35 475 600 735]{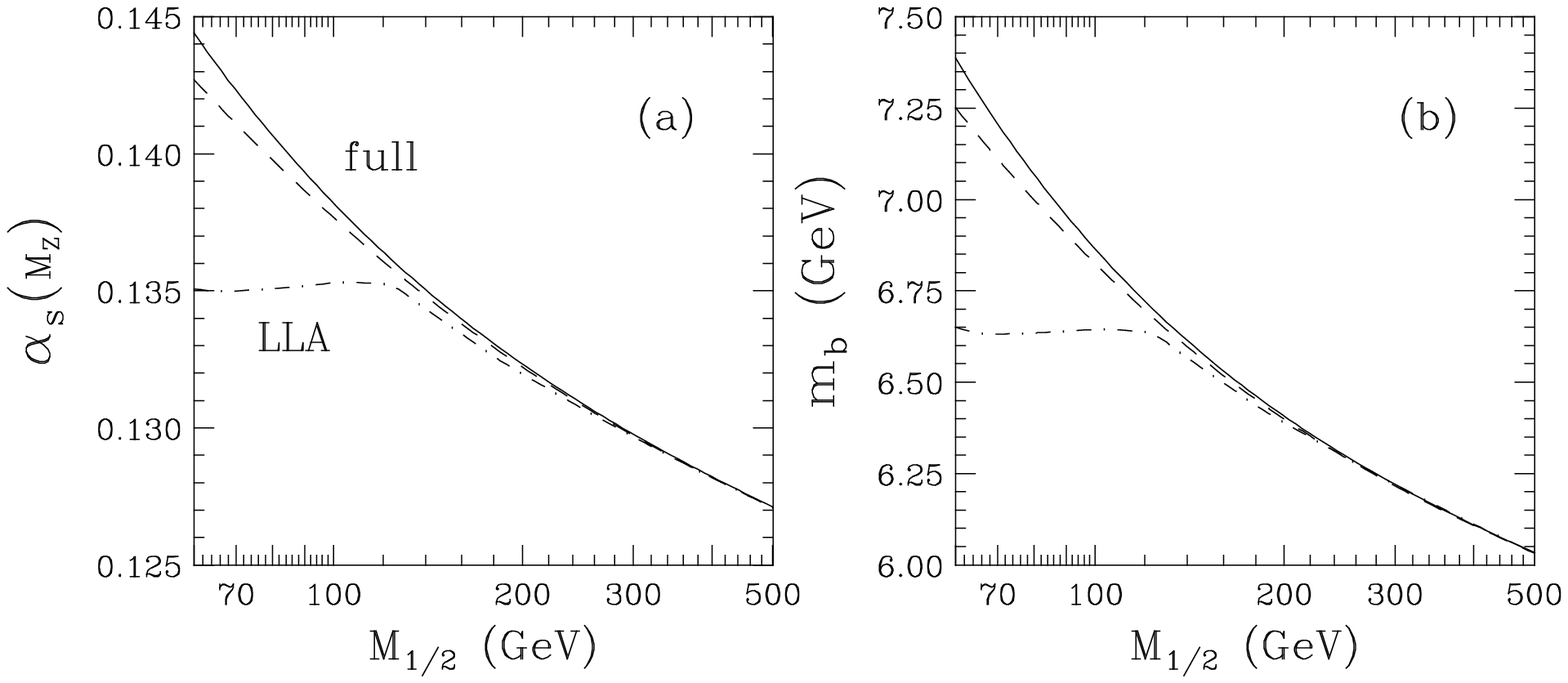}
\begin{center}
\parbox{5.5in}{
\caption[]{\footnotesize Comparison of (a) $\alpha_s$
and (b) $m_b$ in the leading logarithm
approximation (LLA) versus the full one-loop calculation, for $M_0=60$
GeV, $\tan\beta=2, \ m_t=175$ GeV, $A_0=0$, and $\mu>0$.
The dashed line shows the result if
the non-universal corrections are neglected.}}
\end{center}
\end{figure}

In what follows, we have converted the strong coupling
to the \ms scheme so that by $\alpha_s$ we refer to the
standard \ms value evaluated at the scale $M_Z$.
By $m_b$ we refer to the bottom-quark pole mass.

\section{Small $\tan\beta$}

As a reference point, we show in Fig.~3 contours of
$m_b$ and $\alpha_s$ in the $M_0,\ M_{1/2}$ plane, with no GUT
thresholds, $\tan\beta=2$, $m_t=175$ GeV, and $A_0$=0.
We confine our attention to the region of the theory
which is more natural, i.e. to the region where the
superpartner masses are less than about 1 TeV.
We find the strong coupling is large ($\alpha_s>0.127$) compared to
the PDG value \cite{pdg} $\alpha_s=0.117\pm0.005$. Similarly,
the bottom mass is quite large ($m_b>6$ GeV), far outside the preferred
region which we take to be $4.7 < m_b < 5.2$ GeV.
\begin{figure}[t]
\epsfysize=3in
\epsffile[30 468 600 728]{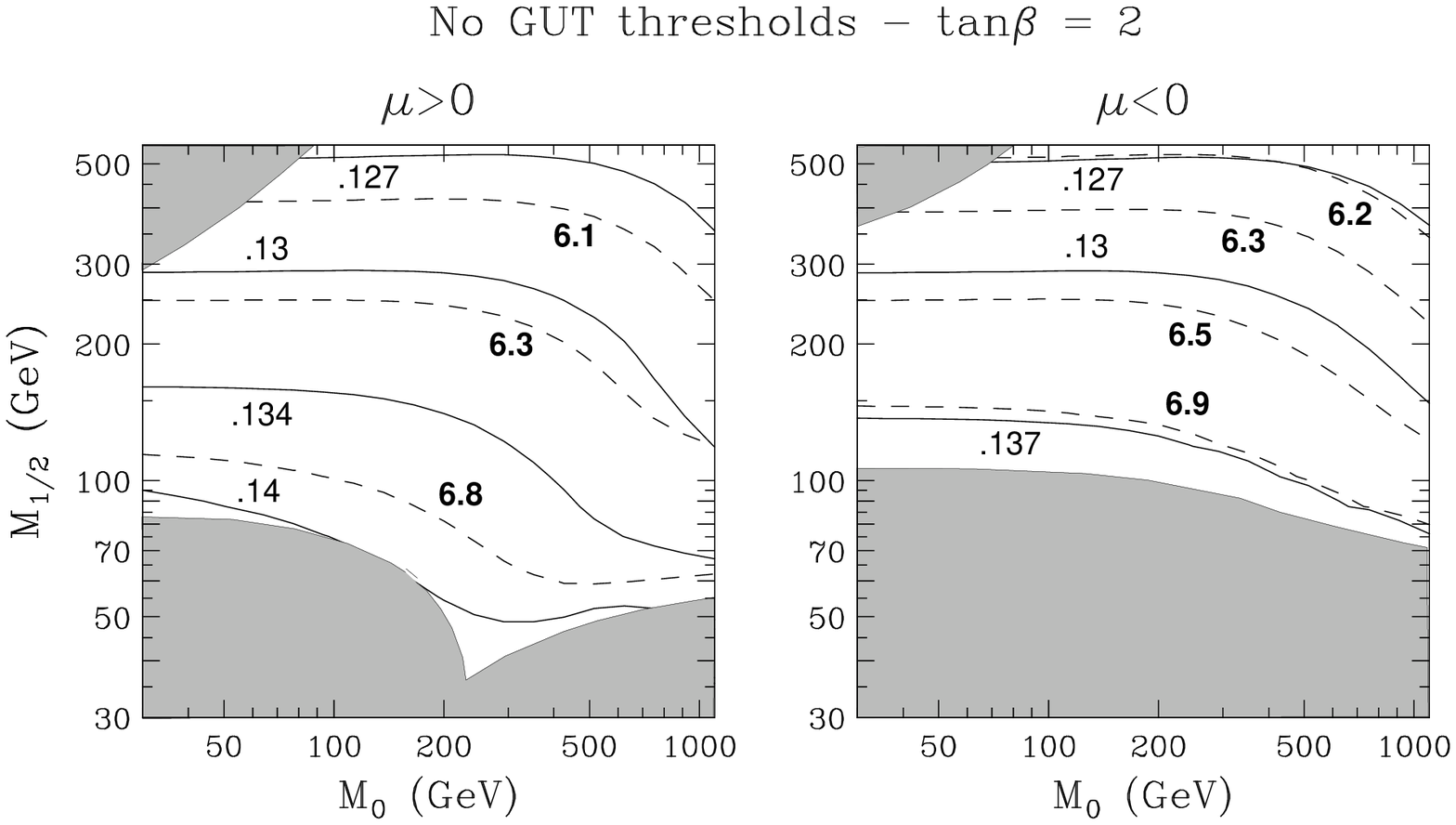}
\begin{center}
\parbox{5.5in}{
\caption[]{\footnotesize Contours of $\alpha_s$ (solid lines)
and $m_b$ (dashed) in the $M_0$, $M_{1/2}$
plane, with $\tan\beta=2$, $m_t=175$ GeV, and $A_0=0$.
The upper left hand corners are excluded on cosmological grounds
(charged LSP) and the lower shaded regions are excluded by
particle searches. The $m_b$ contours are labeled in GeV.}}
\end{center}
\end{figure}

The experimental uncertainty in the determination of $\alpha_s$ is
primarily due to the uncertainty in determining the electromagnetic
coupling at the $Z$-scale.  We use the value recently determined by
Eidelman and Jegerlehner \cite{Eidelman}.
The one-sigma uncertainty in our input
$\alpha_{EM}(M_Z)$ results in a one-sigma uncertainty
in our output $\alpha_s$ of about $\pm0.001$, and an uncertainty
of typically 0.07 GeV in our bottom mass prediction.
Martin and Zeppenfeld \cite{Martin} and Swartz \cite{Swartz}
have also performed analyses to determine $\alpha_{EM}(M_Z)$.
The central value for $\alpha_s$ increases by about 0.001
if we use the value of $\alpha_{\rm EM}(M_Z)$ as determined by
Martin and Zeppenfeld, and it increases by about 0.002 if
we use the value of $\alpha_{EM}(M_Z)$ from Swartz.

The bottom mass is reduced if we go further into the small
$\tan\beta$ region, due to the large top-quark Yukawa coupling
in the bottom mass renormalization group equation.
We consider the case where the top Yukawa is as large as possible;
we set $\lambda_t(M_{\rm GUT})=3$, which is on the verge of the
nonperturbative regime. In this case we will obtain the smallest
possible bottom mass. As seen in Fig.~4,
for $m_t=180$ GeV the bottom mass is less than 5.2 GeV
and the strong coupling is less than 0.127 only if
the squark masses are in the TeV region.
\begin{figure}[t]
\epsfysize=3.5in
\epsffile[25 398 600 718]{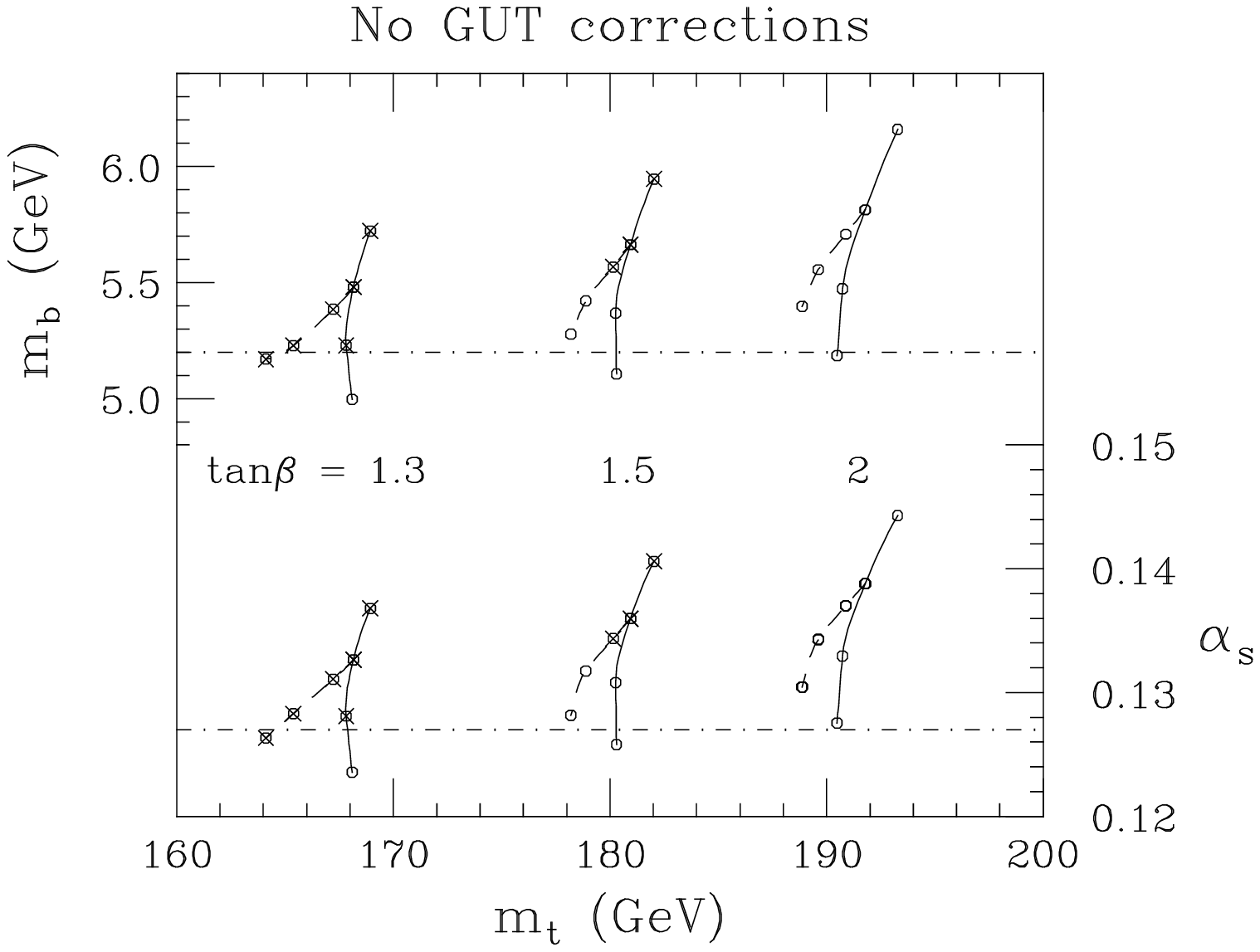}
\begin{center}
\parbox{5.5in}{
\caption[]{\footnotesize The bottom-quark mass and $\alpha_s$
vs. $m_t$ for the case of no GUT-scale thresholds,
for various values of $\tan\beta$, with $A_0=0$, $\mu>0$, and
$\lambda_t(M_{\rm GUT})=3$. The circles on the right (solid) leg in
each pair of lines corresponds to $M_{1/2}$ equal to (from top to
bottom) 60, 100, 200, 500 GeV, with $M_0$ fixed at 100 GeV.
The circles on the left (dashed) leg corresponds to $M_0$ equal to
100, 200, 400 and 1000 GeV, with $M_{1/2}=100$ GeV.
The horizontal dot-dashed lines indicate
$m_b=5.2$ GeV and $\alpha_s=0.127$. The $\times$'s mark points
with one-loop Higgs mass $m_h<60$ GeV.}}
\end{center}
\end{figure}

If we consider particular GUT models, we can determine whether the GUT
threshold corrections to the gauge and Yukawa couplings can help improve
the situation. We parametrize
the GUT threshold corrections by $\varepsilon_g$ and
$\varepsilon_b$, where
$$ g_3(M_{\rm GUT}) = g_{\rm GUT}(M_{\rm GUT})\left(
1+\varepsilon_g\right)\ ,$$
$$ \lambda_b(M_{\rm GUT})=\lambda_\tau(M_{\rm GUT})\left(
1+\varepsilon_b\right)\ ,$$
where $\lambda_b$ and $\lambda_\tau$ are the $b$- and $\tau$-Yukawa
couplings, and $M_{\rm GUT}$ is defined as the scale at which $g_1$
and $g_2$ meet, $g_{\rm GUT}\equiv g_1(M_{\rm GUT})=g_2(M_{\rm GUT})$.
A smaller value of $\alpha_s$ requires $\varepsilon_g<0$.
In the small $\tan\beta$ region the bottom mass is tightly correlated
with the value of the strong coupling.  Hence, setting
$\varepsilon_g<0$ reduces both $\alpha_s$ and $m_b$. In fact,
the bottom mass is an order of magnitude more sensitive to
$\varepsilon_g$ than to $\varepsilon_b$.
In what follows we examine the GUT corrections
in two SU(5) GUT models.
\begin{figure}[t]
\epsfysize=3in
\epsffile[30 468 600 728]{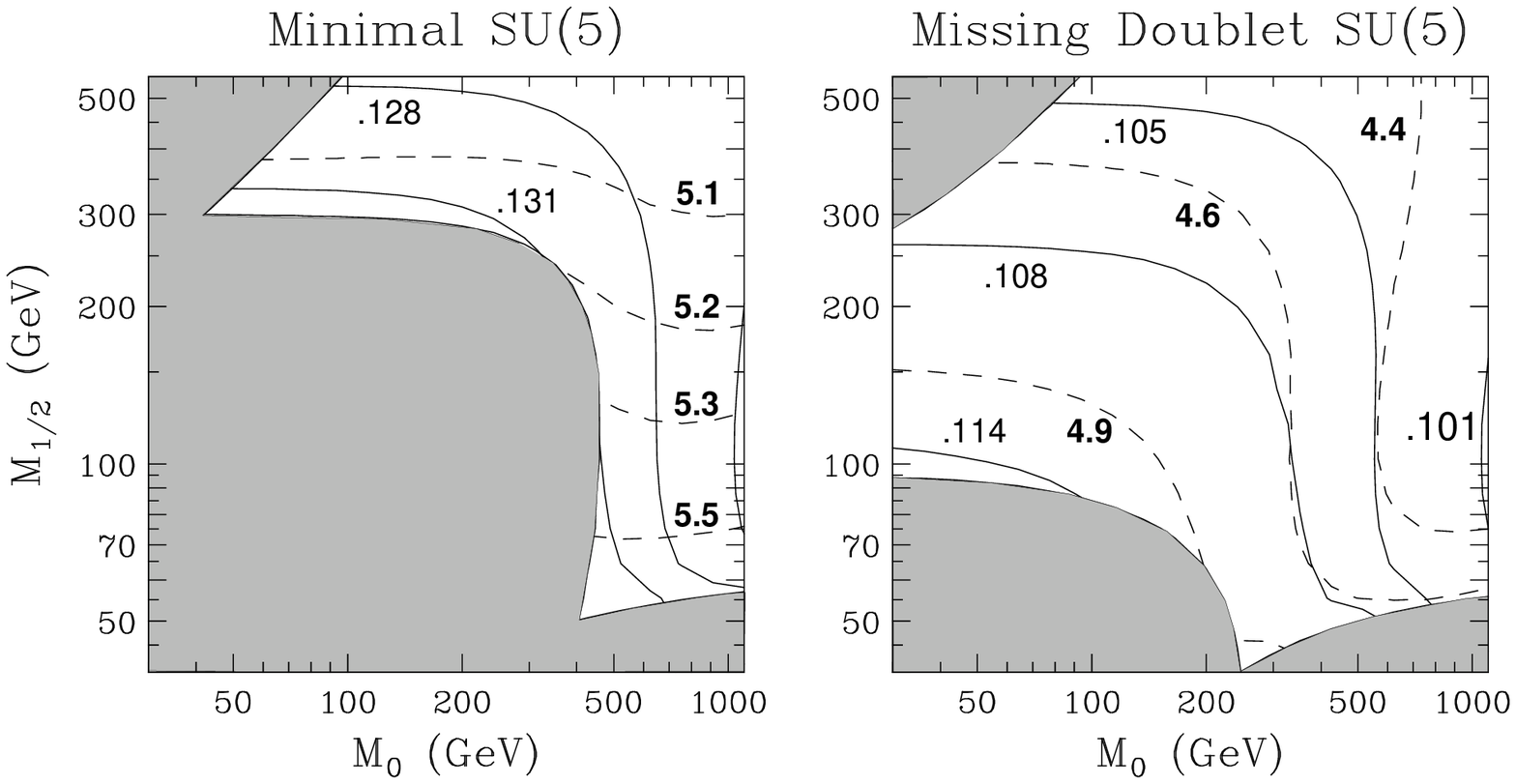}
\begin{center}
\parbox{5.5in}{
\caption[]{\footnotesize Contours of the smallest possible $\alpha_s$
(solid lines) and $m_b$ (dashed) consistent with nucleon decay
in (a) minimal SU(5), and (b) missing doublet SU(5), with
$m_t=175$ GeV, $A_0=0$ and $\mu>0$. In (a) $\lambda_t(M_{\rm GUT})=3$
($\tan\beta\simeq1.4$) and in (b) $\tan\beta=2$.
The shaded regions are phenomenologically excluded.
The $m_b$ contours are labeled in GeV.}}
\end{center}
\end{figure}
\begin{figure}[t]
\epsfysize=3in
\epsffile[30 468 600 728]{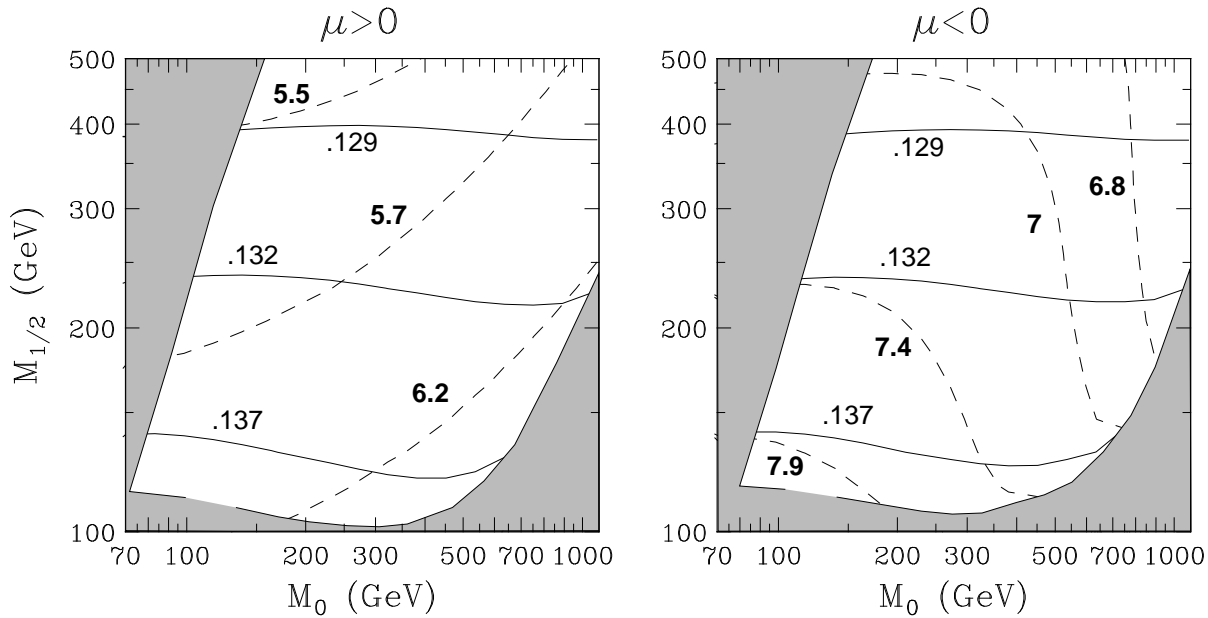}
\begin{center}
\parbox{5.5in}{
\caption[]{\footnotesize Contours of $\alpha_s$ (solid)
and $m_b$ (dashed, labeled in GeV) with no GUT thresholds,
$m_t=175$ GeV, $A_0=0$ and $\tan\beta=30$.}}
\end{center}
\end{figure}
\begin{figure}[t]
\epsfysize=3in
\epsffile[30 468 600 728]{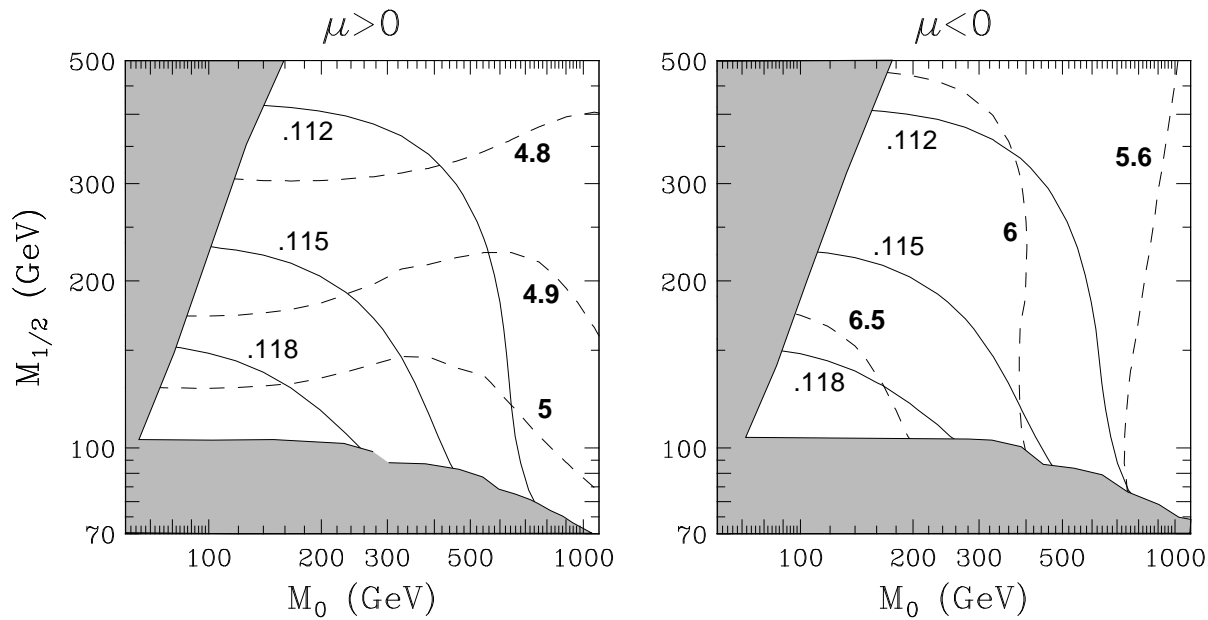}
\begin{center}
\parbox{5.5in}{
\caption[]{\footnotesize Contours of $\alpha_s$ (solid)
and $m_b$ (dashed, labeled in GeV) in missing doublet SU(5), with
$m_t=175$ GeV, $A_0=0$ and $\tan\beta=30$.}}
\end{center}
\end{figure}

In the minimal SU(5) model \cite{minSUfive}, the gauge coupling
threshold correction $\varepsilon_g'$ is given by
 \cite{minSUfive epsg}
\begin{equation}
\varepsilon'_g\ =\ {3g_{\rm GUT}^2\o40\pi^2}\,\log\left(M_{H_3}\o
M_{\rm GUT}\right)\ , \label{epsprime}
\end{equation}
where $M_{H_3}$ is the mass of the color-triplet Higgs particle that mediates
nucleon decay. From this expression, we see that $\varepsilon'_g < 0$ whenever
$M_{H_3} < M_{\rm GUT}$.  However, $M_{H_3}$ is bounded from below by proton
decay experiments. The $M_{H_3}$ mass limit is of the form \cite{Hisano}
$$ M_{H_3} > \M\ {|1+y^{tK}|\over\sin2\beta} f(\tilde w, \tilde d,
\tilde u,\tilde e) $$
where $\M$ is a nuclear matrix element, $y^{tK}$ parametrizes
the amount of third generation mixing, and $f$ is a function
of the wino, squark and slepton masses.

For the conservative choices $\M=0.003$ GeV$^3$
and $|1+y^{tK}|=0.4$ we find that $M_{H_3}^{\rm min}>M_{\rm GUT}$
unless $M_0>500$ GeV and $M_{1/2} \ll M_0$. Thus, in most of the
parameter space, $\varepsilon_g'>0$.
For this reason, in minimal SU(5), $\alpha_s$ is typically even
larger than in the case of no GUT thresholds, as illustrated in
Fig.~5(a). In order to obtain the smallest possible $\alpha_s$ and
$m_b$, we have set $\lambda_t(M_{\rm GUT})=3$ in Fig.~5(a).
Thus we end up with rather small values for $\tan\beta$
($\sim 1.3$-1.6), and the Higgs mass constraint
rules out a large part of parameter space.
Only in the region $M_0 \gg M_{1/2}$, where the proton
decay amplitude is suppressed, is the strong coupling
reduced relative to the case with no GUT thresholds.
The smallest
value of $\alpha_s$ occurs in this region, a somewhat large value of
0.124. The bottom-quark mass is similarly
on the high side of the preferred region.
We have applied the most favorable Yukawa correction $\varepsilon_b$
given in Wright~\cite{wright}, subject to Yukawa coupling perturbativity
constraints (see Bagger et al.~\cite{bmp}).

The missing-doublet model is an alternative SU(5) theory in which
the heavy color-triplet Higgs particles are split naturally
from the light Higgs doublets \cite{missing-doublet}.  In this
model the GUT gauge threshold correction is given by \cite{Yamada}
\begin{equation}
\varepsilon_g^{\prime\prime}\ =
\ {3g_{\rm GUT}^2\o40\pi^2}\,\Biggl\{\log\left(M_{H_3}^{\rm eff}
\o M_{\rm GUT}\right) - {25\o2}\log5 + 15\log2\Biggr\}\ \simeq
\ \varepsilon'_g - 4\%\ .
\label{mdmodel}
\end{equation}
Thus, for fixed $M_{H_3}$, the missing-doublet model has the same
threshold correction as the minimal SU(5) model, minus 4\%.  In
eq.~(\ref{mdmodel}), $M_{H_3}^{\rm eff}$ is the effective mass that enters
into the proton decay amplitude, so the bounds on $M_{H_3}$ in the minimal
SU(5) model also apply to $M_{H_3}^{\rm eff}$ in the missing-doublet
model.

The large negative correction in eq.~(\ref{mdmodel}) is due to the mass
splitting in the {\bf 75} representation, and gives rise to much smaller
values for $\alpha_s$ and $m_b$.  This is illustrated in
Fig.~5(b), where we show contours of $\alpha_s$ and $m_b$
in the $M_0,\ M_{1/2}$
plane, with $M_{H_3}^{\rm eff}=M_{H_3}^{\rm min}$, at $\tan\beta=2$.
We find (even without going into the far infrared top Yukawa fixed point
region) values of both the strong coupling and the bottom mass near
their central values.


\section{Large $\tan\beta$}

Again, as a reference point, we show in Fig.~6 the strong coupling
and bottom mass
prediction with no GUT corrections, for $A_0=0,\ m_t=175$ GeV,
at $\tan\beta=30$. The $\alpha_s$ predictions are not significantly
different from the small $\tan\beta$ case. The bottom mass
is significantly influenced by the large finite corrections~\cite{mb},
$${\Delta m_b\over m_b}\sim -{\mu\tan\beta\over16\pi^2m_{\tilde q}^2}
\left({8\over3}g_3^2 m_{\tilde g} + \lambda_t^2 A_t\right)\ ,$$
so much so, that for the case $\mu<0$ the bottom mass
is hopelessly large. However, these help to reduce $m_b$
when $\mu>0$. Even so, we see in Fig. 6(a) that $m_b > 5.5$ GeV.
GUT threshold corrections are needed to reduce both $\alpha_s$ and
$m_b$ further.

In minimal SU(5) the proton decay rate is enhanced at large $\tan\beta$,
so the triplet Higgs mass $M_{H_3}$ is forced to be quite large.
Hence, we have a large and positive GUT correction $\varepsilon_g'$
(Eq. (\ref{epsprime})) which forces the strong coupling
to unacceptably large values (\roughly{>}0.14).
Hence we conclude that minimal SU(5) is ruled out at large $\tan\beta$.

In the missing doublet model the large triplet Higgs mass
correction is adequately compensated by the constant $-4\%$ correction
of Eq. (\ref{mdmodel}). As shown in Fig. 7, this cancellation results in
near central values for $\alpha_s$ and $m_b$.
However, $\mu>0$ is clearly
required when $\tan\beta$ becomes large, as the large finite corrections
have the wrong sign in the $\mu<0$ case, yielding values of $m_b$
which are unacceptably large.

\section{Conclusion}
\vspace{-.25cm}

In this talk we have presented results from a complete
calculation of the one-loop corrections to the masses, gauge, and
Yukawa couplings in the MSSM. We have seen that such a calculation
allows us to reliably investigate various unified models to
determine whether they are compatible with current experimental data.
In particular, we found that the finite SUSY corrections, which are neglected
in the leading logarithm approximation, can substantially increase the
prediction for $\alpha_s$ and $m_b$ when some of the SUSY partner
masses are lighter than or of order $M_Z$.

For small $\tan\beta$, we found that in the minimal SU(5) model,
$\alpha_s$ was somewhat large ($\alpha_s>0.124$ with $m_{\tilde q}<1$ TeV)
and $m_b$ was larger than 5 GeV. The missing doublet model
gave much lower values of $\alpha_s$ and $m_b$, near their central
values.

For large $\tan\beta$, the minimal SU(5) model GUT threshold correction
became quite large and positive, and this resulted in unacceptably large
values of the strong coupling ($\alpha_s\roughly{>}0.14$).
The missing doublet model,
on the other hand, had no trouble in accomodating the central values of
the strong coupling and $m_b$. The values of $m_b$ were largely
determined by important finite corrections, and we required $\mu>0$ in order
for these corrections to result in acceptably small values of the
bottom-quark mass.

\section{Acknowledgements}
I would like to thank my collaborators Jonathan Bagger and Konstantin
Matchev.  This work was supported
by the U.S. National Science Foundation under grant NSF-PHY-9404057.

\end{document}